# Absence of superconductivity in bulk $Nd_{1-x}Sr_xNiO_2$

Qing Li[#], Chengping He[#], Jin Si, Xiyu Zhu[*], Yue Zhang, Hai-Hu Wen[*]

National Laboratory of Solid State Microstructures and Department of Physics, Center for Superconducting Physics and Materials, Collaborative Innovation Center for Advanced Microstructures, Nanjing University, Nanjing 210093, China

**Recently superconductivity at 9 - 15 K was discovered in an infinite-layer nickelate ($Nd_{0.8}Sr_{0.2}NiO_2$ films), which has received enormous attention. Since the $Ni^{1+}$ ionic state in $NdNiO_2$ may have the $3d^9$ outer-shell electronic orbit which resembles that of the cuprates, it is very curious to know whether superconductivity discovered here has similar mechanism as that in the cuprates. By using a three-step method, we successfully synthesize the bulk samples of $Nd_{1-x}Sr_xNiO_2$ ($x$ = 0, 0.2, 0.4). The X-ray diffractions reveal that all the samples contain mainly the infinite layer phase of 112 with some amount of segregated Ni. This has also been well proved by the SEM image and the EDS composition analysis. The resistive measurements on the Sr doped samples show insulating behavior without the presence of superconductivity. Temperature dependence of the magnetic moment under high magnetic fields exhibits a Curie-Weiss law feature with the paramagnetic moment of about 2 $\mu_B$/f.u.. By applying pressure on $Nd_{0.8}Sr_{0.2}NiO_2$ up to about 50.2 GPa, we find that the strong insulating behavior at ambient pressure is**

**significantly suppressed, but superconductivity has not been observed either. Since the lattice constants derived from our XRD data are very close to those of the reported superconducting films, we argue that the superconductivity in the reported film may not originate from the expected $Nd_{0.8}Sr_{0.2}NiO_2$, but arise from the interface or the stress effect.**

Since the discovery of high critical temperature superconductivity (HTS) in cuprates in 1986 [1], there are plenty of experimental and theoretical studies to explore the intrinsic mechanism for superconductivity [2–5]. There is a general agreement that the parent compound of cuprates like $La_2CuO_4$ is a Mott insulator with a charge transfer gap and long-range antiferromagnetic (AF) order [6]. With chemical doping, the long-range AF order will be suppressed at a hole doping level (p ≈ 0.02) and *d*-wave superconductivity emerges at a higher doping level (p ≥ 0.05) [7–9]. After the efforts more than three decades, some common features have been observed, although the intrinsic pairing mechanism of HTS remains unresolved yet. These include two-dimensional electronic structure and coexistence with an AF order or spin fluctuations, all these also occur in most iron-based [10] and heavy fermion superconductors [11]. Moreover, in cuprates, it is also important that the spin S = 1/2 magnetic moment from $3d^9$ electrons form the basic structure of the AF order. One intuitive way to explore the pairing mechanism of cuprates is to find additional *high-T*$_C$ superconductors with different transition metals but similar crystal and electronic structure [12, 13]. The infinite-layer materials $R$NiO$_2$ ($R$ = La, Nd) are one of the ideal systems to simulate cuprates. First, the $R$NiO$_2$ compounds have the same crystal structure (*P4/mmm*) as CaCuO$_2$, which is a parent compound of *high-T$_C$* cuprates and can reach a *high-T$_C$* of about 110 K by hole doping [14]. Second, the Ni$^{1+}$(*3d$^9$*) oxidation state in $R$NiO$_2$ is very similar to *3d$^9$* configuration of Cu$^{2+}$ in cuprates. Thus, many theoretical and experimental efforts have been put forward to investigate the $R$NiO$_2$ as a promising candidate of cuprate-like Ni-based superconductor [12, 15–17].

Recently, superconductivity was observed at 9 - 15 K in the strontium doped infinite-layer nickelate thin films of Nd$_{0.8}$Sr$_{0.2}$NiO$_2$ on SrTiO$_3$ substrate [18]. This work has stimulated enormous interests, and plenty of theoretical works have been carried out [19–31]. Among them, Zhang *et al.* propose the parent compound of Nd$_{0.8}$Sr$_{0.2}$NiO$_2$ as a self-doped Mott insulator [26]. This work suggests that the low-density *Nd-5d* conduction electrons couple with the localized *Ni-3d* electrons, which suppresses the long-range AF order and forms Kondo spin singlets at low temperatures. While, Botana and Norman argue that a large ratio of longer-range hopping to near-neighbor hopping is conducive for superconductivity both in cuprates and nickelate [20]. And Ryee *et al.* demonstrate that magnetic two-dimensionality induced by hole doping is the key factor for superconducting Nd$_{1-x}$Sr$_x$NiO$_2$ [25]. Bernardini *et al.* [31] propose a possible difference between cuprates and nickelates based on the computed London penetration depth, and suggest that the latter does not follow the

Uemura plot which holds well in underdoped cuprates. Hirayama *et al.* [30] compare the electronic structure of curates and NdNiO$_2$, and conclude that the Nd layer also forms Fermi pockets. And they also propose some other promising compounds analogues to *high-T*$_C$ cuprates. Several other groups also studied the dominant pairing instability in the framework of *t-J* model [23, 27]. They proposed that superconductivity in nickelate has a *d*-wave symmetry, which is analogous to cuprates. However, since the report of discovering superconductivity in Nd$_{0.8}$Sr$_{0.2}$NiO$_2$ [18] thin films, no other experimental works have been reported up to now.

In this paper, we report the successful synthesis and physical properties of bulk Nd$_{1-x}$Sr$_x$NiO$_2$ (*x* = 0, 0.2, 0.4) samples. By using a three-step reaction process, we prepare bulk Sr-doped NdNiO$_2$ polycrystalline samples successfully with the similar doped composition of Sr as in the reported films. The structural and composition analyses reveal the formation of bulk Nd$_{1-x}$Sr$_x$NiO$_2$. Magnetization measurements of the sample show a Curie-Weiss like feature at high fields (10 kOe and 30 kOe). The resistivity measurements at ambient and high pressures exhibit insulating behavior without the presence of superconductivity. We think the contradictory results compared with that of the reported films [17] may be attributed to the interface or stress effect in films. Furthermore, in our samples, we find a slight deficiency (5 - 9 %) of Ni in the NiO$_2$ plane. It would be interesting to know whether this feature occurs also in the reported films, which may lead to the absence of superconductivity in our bulk samples.

## Results

**Sample characterization.** Figure 1a shows the schematic crystal structures and the transformation from the 113 to 112 phase through the low-temperature topotactic reduction method. Here Nd/Sr, Ni and oxygen atoms are represented by the orange/purple, green and red spheres, respectively. To verify the formation of bulk $Nd_{1-x}Sr_xNiO_2$ phase, we conduct the room temperature XRD measurements with $2\theta$ from 10° to 90° and the Rietveld refinements [32] of $Nd_{1-x}Sr_xNiO_2$ ($x$ = 0.2, 0.4) bulk samples. The results are shown in Fig.1 b,c. The XRD data of undoped $NdNiO_2$ sample is shown in Supplementary Fig. 2. All the samples contain mainly the infinite-layer 112 phase (with mole ratio more than 70%). Some amount of extra Ni appears as disconnected segregations. Some peaks seem to be broad, which may be induced by the random orientation of different grains with different sizes. The crystallographic data obtained from the Rietveld refinement profiles are shown in Table I. The lattice constant of $c$ - axis increases with increasing doping contents. The lattice parameters of our bulk sample (x = 0.2) are $a$ = 3.88 Å, $c$ = 3.34 Å, which are in good agreement with the previous reports on the reported $Nd_{0.8}Sr_{0.2}NiO_2$ thin film [18].

Figure 2 a,b display the scanning electron microscope (SEM) images of $Nd_{1-x}Sr_xNiO_2$ ($x$ = 0.2, 0.4) samples. As we can see, the grains could be clearly separated into two major morphologies, which are characterized by the argenteous crystals (dominant phase) and dark grey areas. The argenteous grains are interconnected, while the dark grey ones are well separated with each other. In order to have a deeper insight of the difference between the two different grains, energy dispersive X-ray spectroscopy (EDS) is used to analyze the element concentration of all three samples. In each sample, we measure more than 20 spots randomly (marked by the red crosses in images of Fig. 2a,b and Supplementary Fig. 2). The results are shown in Fig. 2c-f. As there is uncertainty for the composition of carbon and oxygen, here we only study the contents of Nd, Sr and Ni. The typical EDS patterns of different samples are given in Fig. 2c-d, we find that the argenteous crystals have a composition of $Nd_{1-x}Sr_xNi_yO_2$, while the dark grey grains show a composition of Ni (see Supplementary Fig. 3). Two typical features can be identified for the $Nd_{1-x}Sr_xNi_yO_2$: (1) The ratio of Ni: (Nd+Sr) is close to 1, although a slight deficiency of Ni can be seen from the EDS data, as shown in Fig. 2e. The deficiency of Ni, if existing, should be in the scale of 5 - 9 % in all three samples ($x$ = 0, 0.2, 0.4). The vacancies in nickel sites may cause the distortion of $NiO_2$ plane and hence result in the shrinkage of the lattice constants of *a* and *b*. (2) The doped Sr concentrations in $Nd_{1-x}Sr_xNiO_2$ ($x$ = 0.2, 0.4) samples are close to the nominal values, which indicates the effective doping level of Sr in our samples. In Fig. 2f, we show the proportion distribution of Sr concentrations, more than half of the grains reach the nominal concentration of Sr. Therefore, we believe our samples contain the right phase of $Nd_{1-x}Sr_xNiO_2$ (x = 0.2, 0.4). The sensitivity of the SQUID instrument should be sufficient to detect the superconducting signal if these grains are superconductive.

**Magnetic and Electrical transport properties.** Figure 3a-c show the temperature dependence of magnetic moment (*M-T*) curves at different external magnetic fields for $Nd_{1-x}Sr_xNiO_2$ (x = 0, 0.2, 0.4) samples. The magnetic moment of all samples at high fields (*H* = 10 kOe and 30 kOe) increases with decreasing temperature in the whole temperature range, exhibiting approximately linear behavior at high temperature and a Curie-Weiss (*C-W*) like enhancement at low temperature. It should be noted that the low field (*H* = 10 Oe) magnetization of all three samples show a drop of magnetization at about 5 K, together with an obvious irreversibility between the ZFC and FC curves (see Supplementary Fig. 4). Such abnormal behavior at low field magnetization curve may indicate the presence of spin glassy state, probably due to magnetic impurities in compounds. To get a deeper understanding of magnetization behavior, we measure the magnetization hysteresis (*M-H*) curves of the $Nd_{0.8}Sr_{0.2}NiO_2$ sample at different temperatures and show the data in Fig. 3d. We can safely exclude the possibility that the anomalous drop of magnetization at about 5 K is caused by superconductivity, because no superconducting like *M-H* curve is observed at 3 K (inset of Fig. 3d). The *M-H* curves show the presence of paramagnetic background with a ferromagnetic component, and the paramagnetic component is decreasing as the temperature increases. We think the ferromagnetic component is originated from the segregated phase of Ni. And the paramagnetic behavior is attributed to the Nd$_{1-x}$Sr$_x$NiO$_2$ phase. We have tried to fit the high field *M-T* curves in low temperature region by the *C-W* law $\chi = \chi_0 + \frac{C}{T+T_\theta}$. The fitting results are given in Fig. 3e, as shown by the red solid line, indicating a good quality of *C-W* fitting. Through the *C-W* fitting, we derive the effective magnetic moments μ$_{eff}$ of Nd$_{0.8}$Sr$_{0.2}$NiO$_2$ and Nd$_{0.6}$Sr$_{0.4}$NiO$_2$, which are about 2.32 μ$_B$/f.u. and 2.03 μ$_B$/f.u., respectively. It should be noted that, before doing the *C-W* fitting, we need to remove the ferromagnetic background of nickel from the *M-T* curves by subtracting the curves of 1 T from those of 3 T. This is valid since the magnetization of Ni gets already saturated at the magnetic fields of 1 T and 3 T, as shown in Supplementary Fig.5. The subtracted paramagnetic magnetic values $\chi_P = (M_{3T} - M_{1T})/\Delta H$ is thus taken as the pure signal from the paramagnetic term. More details are given in Supplementary material.

Electrical transport measurements are carried out at ambient pressure in the temperature range from 2 to 300 K. Fig. 4a shows the comparison of the temperature dependence of resistivity for Nd$_{1-x}$Sr$_x$NiO$_2$ (x = 0, 0.2, 0.4) samples. It can be clearly seen that the resistivity of the three samples all show insulating behavior. To understand the underlying physics of the electrical transport behavior in present materials, in Fig. 4b we present the $\rho$ (T) curve within the frame of variable range hopping (*VRH*) model [33] described as $\rho =$

$\rho_0 exp(T_0/T)^{-1/4}$, we can see that in a short period of temperature, it gives a linear relationship as highlighted by the red line, while the global fitting is not good. Also the band-gap model ($\ln\rho \propto 1/T$, shown in the bottom-right inset) and resistivity versus $\log 1/T$ (shown in the top-left inset) are plotted, but all failed to fit the data at ambient pressure, suggesting that the insulating is not originated from the band gap and should possess by itself some exotic scattering reasons. Fig. 4c,d show the magnetic field dependent resistivity measured at different temperatures ($T$ = 2, 5, 10, 20, 50 K). A sizable negative magnetoresistance $\Delta\rho/\rho_0 = (\rho_H - \rho_0)/\rho_0$ (about -10% at 50 kOe) is observed at 2 K. The negative magnetoresistance (*MR*) at low temperature decreases with the increase of temperature, then the *MR* shows the crossover from a negative to a small positive magnetoresistance behavior above 20 K. The negative *MR* in correlated oxides may be attributed to various mechanisms such as hopping conduction, magnetic scattering or de-localization effect. As for the positive *MR*, researchers have proposed that the exchange correlation in different hopping sites (variable range hopping type conduction) could give rise to positive *MR*[34, 35]. Owing to the nickel segregations in our samples, the origin of negative and positive *MR* observed in the present compounds is not clear and more work deserve to be done.

**Resistivity under high pressure.** As shown above, an obvious conclusion drawn from the $\rho$ (*T*) curves at ambient pressure is the insulating behavior. One can see that not only the magnitude of resistivity, but also the insulating feature are strongly suppressed by doping more Sr. In order to induce a metallic or even a superconducting state, we also conduct high pressure electrical resistance measurement. The temperature dependence of resistivity at various pressures in $Nd_{0.8}Sr_{0.2}NiO_2$ and $Nd_{0.6}Sr_{0.4}NiO_2$ samples are shown in Fig. 5a and Supplementary Fig. 6. As discussed above, $\rho(T)$ curves of $Nd_{1-x}Sr_xNiO_2$ exhibit an insulating behavior at ambient pressure. With application of pressure, the overall magnitude of the resistivity is continuously reduced. The insulating behavior is significantly weakened. The ratio $\rho_{2K}/\rho_{300K}$ of $Nd_{0.8}Sr_{0.2}NiO_2$ sample is about 350 with the external pressure of 3.2 GPa. With increasing pressure, the magnitude of $\rho_{2K}/\rho_{300K}$ decreases progressively. At the highest pressure of 50.2 GPa, the value of $\rho_{2K}/\rho_{300K}$ has dropped to 6, however the $\rho(T)$ curve still exhibits a semiconducting-like feature in whole temperature region. Then we try to fit the electrical transport behavior of $Nd_{0.8}Sr_{0.2}NiO_2$ sample at the highest pressure and show the fitting in Fig. 5b. The fitting curves of band gap model (the top-left inset) and *VRH* model (the bottom-right inset) are given. The results show that both the band gap model and the *VRH* model fail to fit the data. Surprisingly, the $\rho$ versus $\log(1/T)$ curve at the highest pressure ($P$ = 50.2 GPa) becomes roughly linear, as highlighted by the red linear line in a temperature range between 3.5 K to 23 K. This strange $\rho \propto \log(1/T)$ behavior has been reported in some correlated oxides such as under-doped and over-doped cuprates or vanadium oxide $V_2Se_2O$, which may result from the electron correlation effect in correlated oxides with 3$d$ transition metals [36–38].

## Discussions

After a systematic analysis of the data, a main experimental finding in bulk infinite layer nickelates Nd$_{1-x}$Sr$_x$NiO$_2$ (x = 0, 0.2, 0.4) is the insulating behavior, which is in contrast to the metallic state and the superconductivity below 9 - 15 K in Nd$_{0.8}$Sr$_{0.2}$NiO$_2$ thin film [18]. It is certainly very important to know whether the insulating behavior in the bulk Nd$_{1-x}$Sr$_x$NiO$_2$ is intrinsic. Based on the nice fitting to the XRD data and composition analysis on the grains of the samples, we can safely conclude that the main phase in the samples is definitely Nd$_{1-x}$Sr$_x$NiO$_2$ (x = 0.2, 0.4) with only some disconnected segregation of Ni. From previous literatures, we find that nickel is a metal with the resistivity of about 7.2 mΩ-cm at room temperature [39], so it would be a metallic behavior if the segregated grains of nickel form a conductive network. However, from our SEM image, the segregated nickel grains are well separated with each other. The behavior of resistance both in ambient and high-pressure indicates that the insulating behavior is robust in our samples. Thus, the existence of disconnected nickel regions in samples may only affect the absolute value of resistivity, but not gives rise to the intrinsic insulating behavior. In principle, high pressure can compress the cell volume and reduce the lattice parameters of compound, resulting in insulator-metal transitions and even superconductivity [40, 41]. The clear suppression of insulating behavior in our high-pressure study may be induced by the modification of the bands, or the weakening of the correlation effect, which lead to an enhanced effective density of states at the Fermi energy.

Since we do not find superconductivity in our bulk samples with the same structure and close lattice constants as the reported films [18], we would like to suggest several reasons to interpret this discrepancy. Firstly, the superconductivity in $Nd_{0.8}Sr_{0.2}NiO_2$ films may arise from the interface or the stress effect. In the interface region, both the electron band and the doping level would be strongly modified, which could lead to superconductivity. The second reason may be the slight Ni deficiency (5 - 9 %) in grains of $Nd_{1-x}Sr_xNiO_2$ of our samples, which may cause stronger buckling of the $NiO_2$ planes and lead to a strong localization or scattering of charges. If this is the case, we need to make samples without Ni deficiency. It would be interesting to know whether the reported films also have Ni deficiency.

To summarize, we have successfully synthesized bulk $Nd_{1-x}Sr_xNiO_2$ (x =0, 0.2, 0.4) samples and performed comprehensive studies of the physical properties. From structure and composition analyses by XRD and EDS, the tetragonal $Nd_{1-x}Sr_xNiO_2$ (x =0, 0.2, 0.4) phase is established and the strontium content is sufficient in most grains. The electrical transport and magnetization measurements reveal an insulating behavior and a Curie-Weiss like feature with a ferromagnetic background caused by nickel segregations. Electrical transport measurements under high pressure show that the insulating behavior can be effectively suppressed but superconductivity is still not observed.

## Methods

I. **Sample growth**

To synthesize polycrystalline samples of $Nd_{1-x}Sr_xNiO_2$ (x = 0, 0.2, 0.4), we first prepare polycrystalline $Nd_{2-2x}Sr_{2x}NiO_4$ (x =0, 0.2, 0.4) samples by the solid-state reaction of $Nd_2O_3$, NiO, and SrO at 1200 °C for 24h. And then we use precursor $Nd_{2-2x}Sr_{2x}NiO_4$, NiO, and $KClO_4$ to synthesize the $Nd_{1-x}Sr_xNiO_3$ (x =0, 0.2, 0.4) with high pressure and high temperature. Here $KClO_4$ is used as an excess oxygen source. Then the mixture is pressed into a pellet and sealed in a gold capsule. This procedure is done in a glove box with oxygen and water concentrations less than 0.1 ppm. The gold capsule is placed in a *BN* capsule and heated up to 1000 °C and stayed for 2 h at this temperature under a pressure of 2 GPa. The resultant compound is the perovskite $Nd_{1-x}Sr_xNiO_3$ phase which is verified by powder X-ray diffraction at room temperature (see Supplementary Fig. 1). Next, the samples of $Nd_{1-x}Sr_xNiO_2$ are obtained by reacting $Nd_{1-x}Sr_xNiO_3$ (x = 0, 0.2, 0.4) with $CaH_2$, which is similar to the previous report [42, 43].

II. **Physical properties measurements**

The crystal structures of the prepared samples are determined by powder X-ray diffraction (XRD) (Bruker *D8 Advance*) using Cu-*Kα* radiation at room temperature with a scanning step of 0.01° and $2\theta$ from 10° to 90°. The Rietveld refinements are done based on the *TOPAS4.2* software [32, 44]. The SEM photograph of the polycrystalline and the energy dispersive X-ray microanalysis spectrum are done by Phenom Pro*X* (Phenom). The measurements are performed at an accelerating voltage of 15 kV. The DC magnetization is measured with a SQUID based on the vibrating sample technique (*SQUID-VSM 7T*, Quantum Design). The electrical resistances at ambient pressure are measured by the standard four-probe method using the physical property measurement system (*PPMS 16T*, Quantum Design). The high pressure resistivity measurements are performed by using the diamond avail cell (*cryo*DACPPMS, Almax easyLab) [45].

## Acknowledgements

This work was supported by the National Key R&D Program of China (Grant No. 2016YFA0300401 and 2016YFA0401704), National Natural Science Foundation of China (Grant No. A0402/11534005 and A0402/11674164), and the Strategic Priority Research Program of Chinese Academy of Sciences (Grant No. XDB25000000).

## Author Contributions

The samples were grown by Q.L., C.P.H, Y.Z., and X.Y.Z. The physical properties measurements were conducted by Q.L., and J.S. The structure and composition analyses were done by Q.L., C.P.H. and X.Y.Z. H.-H.W., X.Y.Z. and Q.L. wrote the manuscript with the supplementary by others. All authors have discussed the results and the interpretations.

## Author Information

The authors declare no competing financial interests. The authors marked with # contribute equally to the work. Correspondence and requests for materials should be addressed to X.Y.Z. and H-H.W. (zhuxiyu@nju.edu.cn, hhwen@nju.edu.cn,).

**Table I**

| Compound | Space group | a, b (Å) | c (Å) | V (Å) | $R_{wp}$(%) | $R_p$(%) | GOF |
|---|---|---|---|---|---|---|---|
| $NdNiO_2$ | *P4/mmm* | 3.914(2) | 3.239(6) | 49.63(5) | 5.23 | 4.00 | 1.28 |
| $Nd_{0.8}Sr_{0.2}NiO_2$ | *P4/mmm* | 3.9138(9) | 3.3303(8) | 51.01(6) | 4.16 | 3.30 | 1.09 |
| $Nd_{0.6}Sr_{0.4}NiO_2$ | *P4/mmm* | 3.8850(6) | 3.3425(1) | 50.44(9) | 4.37 | 3.38 | 1.33 |
| $Nd_{0.8}Sr_{0.2}NiO_2$ thin films [18] | *P4/mmm* | 3.91 | 3.34-3.38 | -- | -- | -- | -- |

**Table 1 | Crystallographic parameters obtained from the Rietveld refinements of $Nd_{1-x}Sr_xNiO_2$ ($x$ = 0, 0.2, 0.4) at room temperature and a comparation with superconducting $Nd_{0.8}Sr_{0.2}NiO_2$ thin film [18].**

# Figures and legends

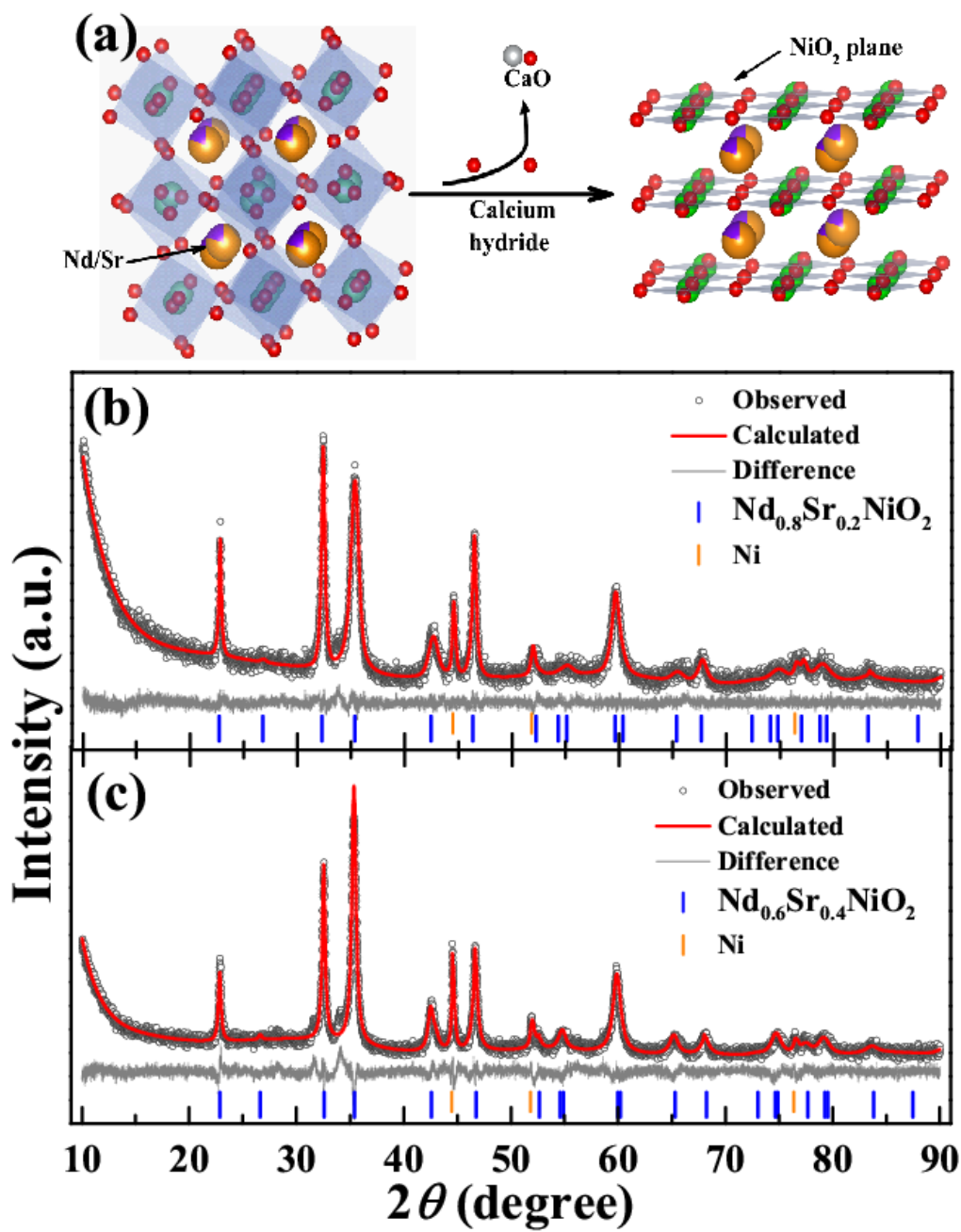

**Figure 1 | Crystal structure and Rietveld refinement analyses. a,** The schematic structure and transformation from perovskite $Nd_{1-x}Sr_xNiO_3$ to infinite-layer $Nd_{1-x}Sr_xNiO_2$ by low-temperature reduction process with $CaH_2$. **b-c,** Powder X-ray diffraction patterns of $Nd_{0.8}Sr_{0.2}NiO_2$ and $Nd_{0.6}Sr_{0.4}NiO_2$ (circles) and Rietveld fitting curves (red lines) to the data.

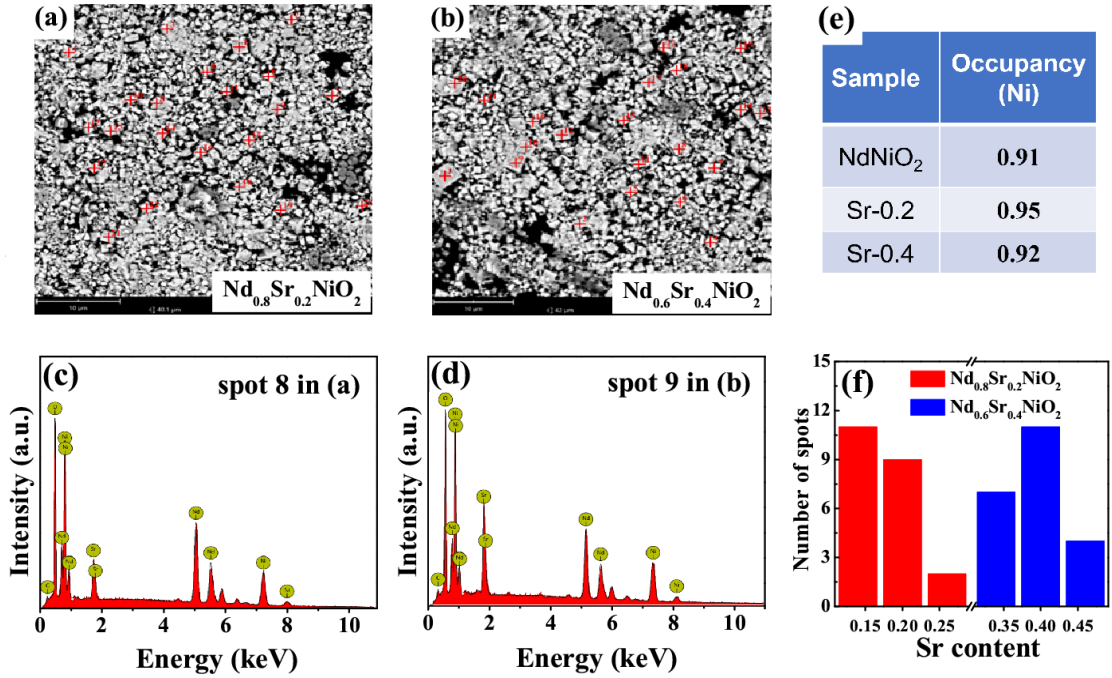

**Figure 2 | Scanning electron micrograph images and energy dispersive spectroscopy analyses. a-b,** SEM images of $Nd_{0.8}Sr_{0.2}NiO_2$ and $Nd_{0.6}Sr_{0.4}NiO_2$ samples. **c-d,** The typical energy dispersive spectroscopy (EDS) of the spot 8 in (a) and spot 9 in (b), respectively. **e,** The occupied ratio of nickel in $Nd_{1-x}Sr_xNiO_2$ ($x$ = 0, 0.2, 0.4) samples. **f,** The proportion distribution of Sr concentration by measuring EDS on different grains of the samples.

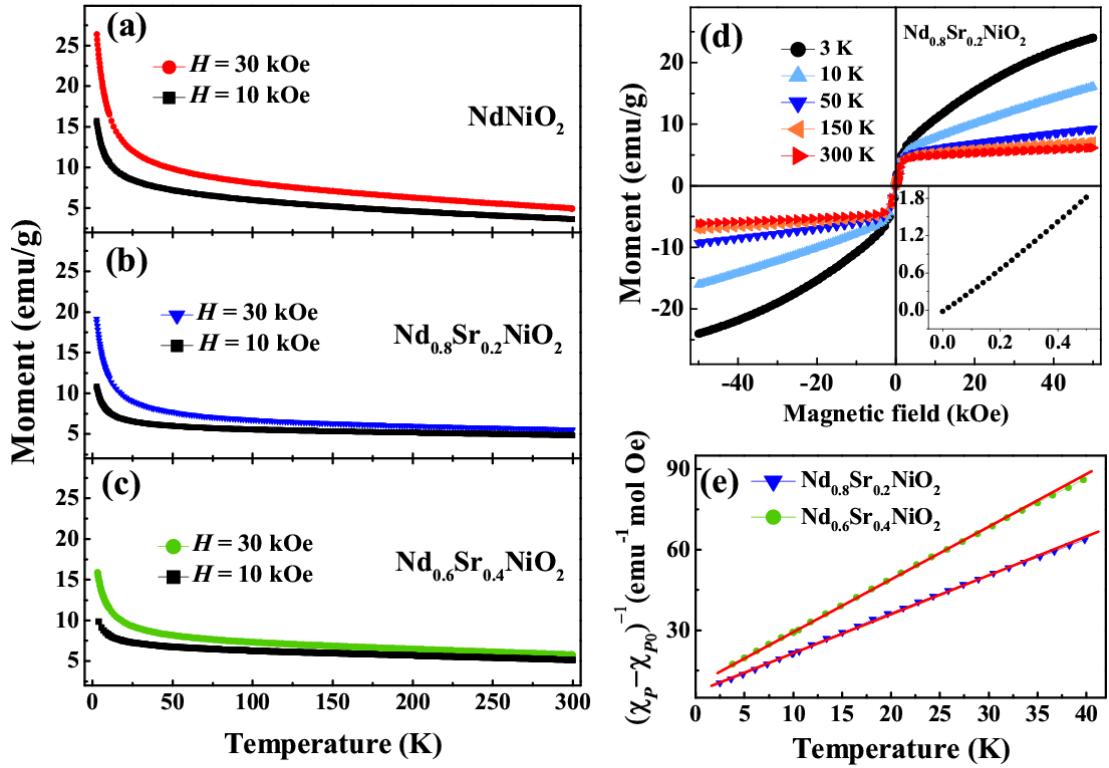

**Figure 3 | Magnetic properties of $Nd_{1-x}Sr_xNiO_2$ ($x$ = 0, 0.2, 0.4) samples. a-c,** Temperature dependence of magnetic moment measured at the fields of 10 kOe and 30 kOe for $NdNiO_2$, $Nd_{0.8}Sr_{0.2}NiO_2$ and $Nd_{0.6}Sr_{0.4}NiO_2$ samples. **d,** Magnetization hysteresis loops for $Nd_{0.8}Sr_{0.2}NiO_2$ sample at different temperatures. Inset shows a linear relation between $M$ with $H$ at low field. **e,** Temperature dependence of $(\chi_P - \chi_{P0})^{-1}$ in the low temperature region. The red line shows a linear relation of $(\chi_P - \chi_{P0})^{-1}$ versus $T$.

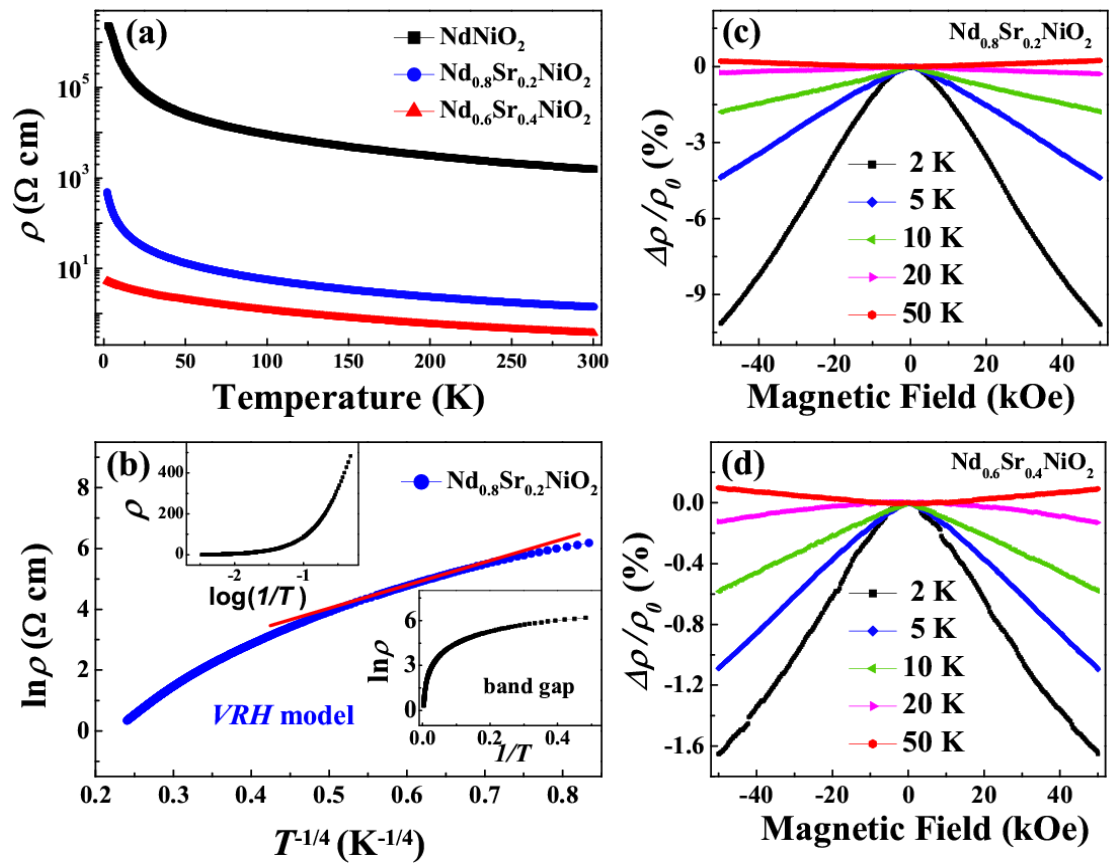

**Figure 4 | Temperature dependent resistivity and magnetoresistance at ambient pressure. a,** Temperature dependence of resistivity for Nd$_{1-x}$Sr$_x$NiO$_2$ ($x$ = 0, 0.2, 0.4) samples. **b,** Temperature dependence of resistivity in ln$\rho$ versus $T^{-1/4}$ for Nd$_{0.8}$Sr$_{0.2}$NiO$_2$. The top inset shows the curve of $\rho$ versus log(1/T) and the bottom inset shows the curve of ln$\rho$ versus 1/$T$ (band gap model) for Nd$_{0.8}$Sr$_{0.2}$NiO$_2$. **c,d,** Field dependence of magnetoresistance $\Delta\rho/\rho_0$ at different temperatures for Nd$_{0.8}$Sr$_{0.2}$NiO$_2$ and Nd$_{0.6}$Sr$_{0.4}$NiO$_2$ samples. Both samples exhibit a reversal from negative magnetoresistance to positive magnetoresistance above 20 K.

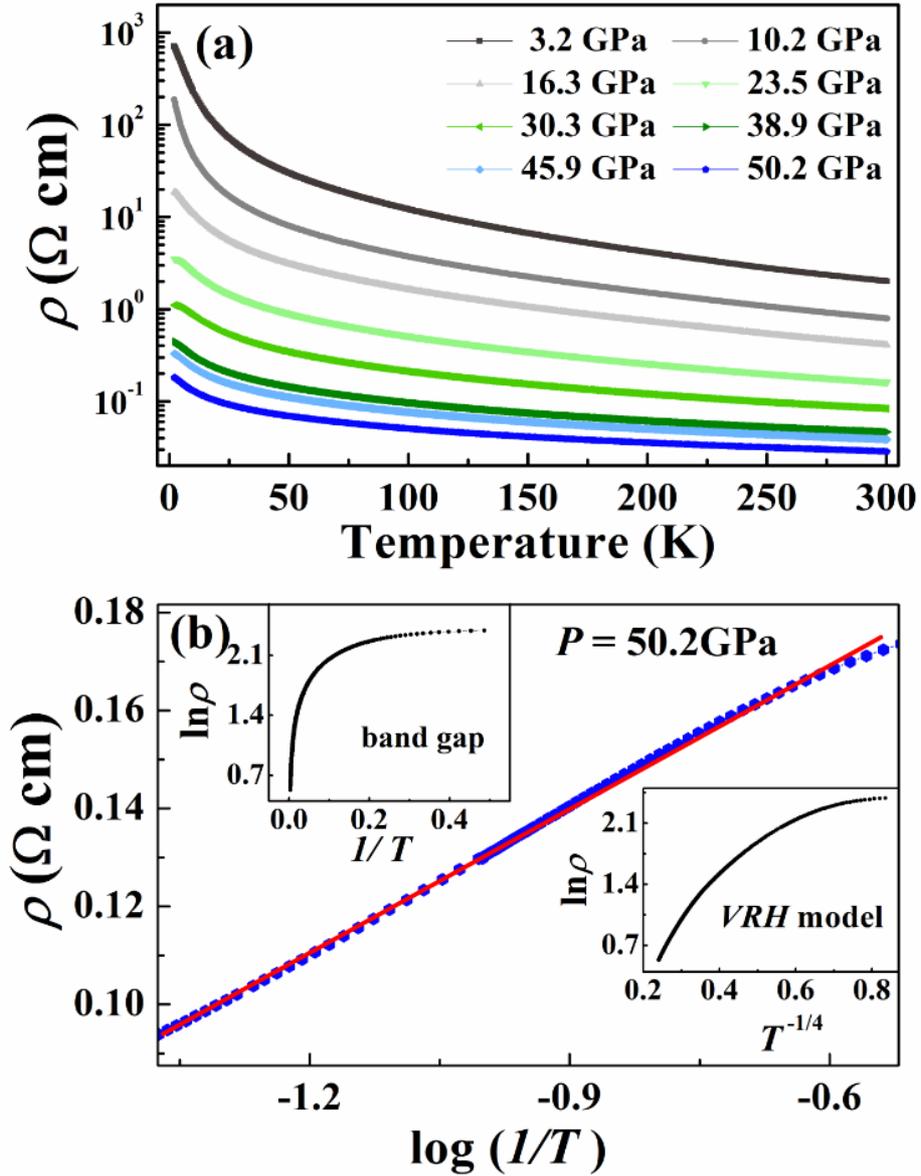

**Figure 5 | Temperature dependent resistivity of $Nd_{0.8}Sr_{0.2}NiO_2$ under pressures. a,** Temperature dependence of resistivity for $Nd_{0.8}Sr_{0.2}NiO_2$ at various pressures. **b,** The $\rho$ versus ln(1/$T$) curve for $Nd_{0.8}Sr_{0.2}NiO_2$ in the temperature range from 3.5 K to 23 K at the pressure of 50.2 GPa together with the corresponding linear fit (red line). The top inset shows the curve of ln$\rho$ versus 1/$T$, corresponding to band gap model. The bottom inset shows the curve of ln$\rho$ versus $T^{-1/4}$, corresponding to the VRH model.

# Supplementary Information

# Absence of superconductivity in bulk $Nd_{1-x}Sr_xNiO_2$

Qing Li, Chengping He, Jin Si, Xiyu Zhu*, Yue Zhang, Hai-Hu Wen*

National Laboratory of Solid State Microstructures and Department of Physics, Center for Superconducting Physics and Materials, Collaborative Innovation Center for Advanced Microstructures, Nanjing University, Nanjing 210093, China

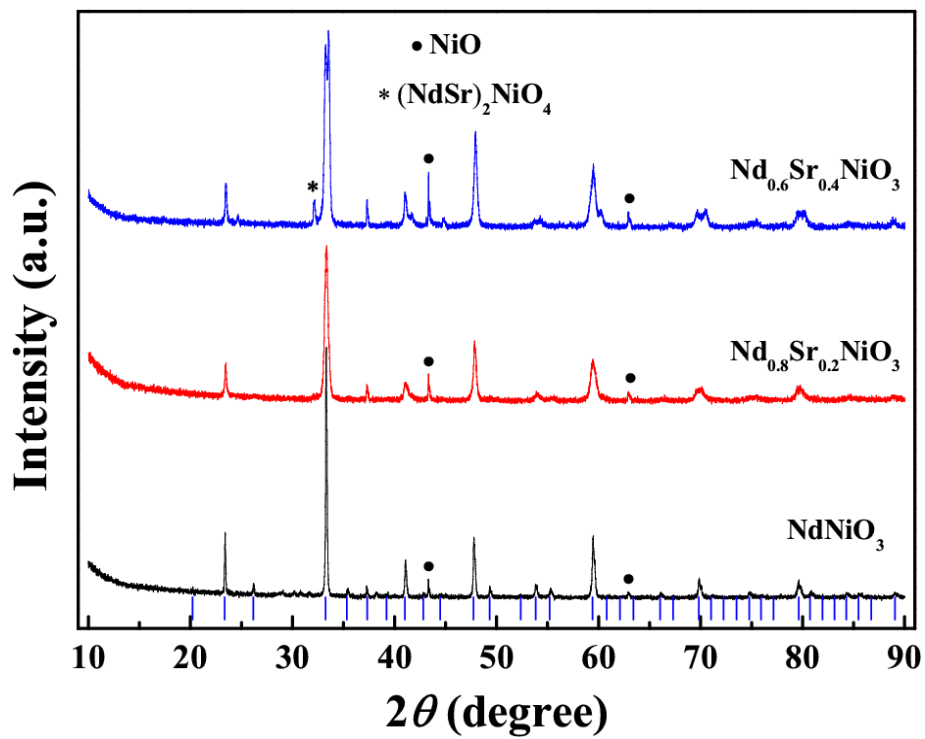

**Supplementary Figure 1 | The X-ray diffraction patterns of $Nd_{1-x}Sr_xNiO_3$ (x =0, 0.2, 0.4) powders.** All three samples contain small amount of NiO impurities marked with black dot (•), and one peak marked with asterisk (∗) in $Nd_{0.6}Sr_{0.4}NiO_3$ may come from the Sr-doped $Nd_2NiO_4$ phase.

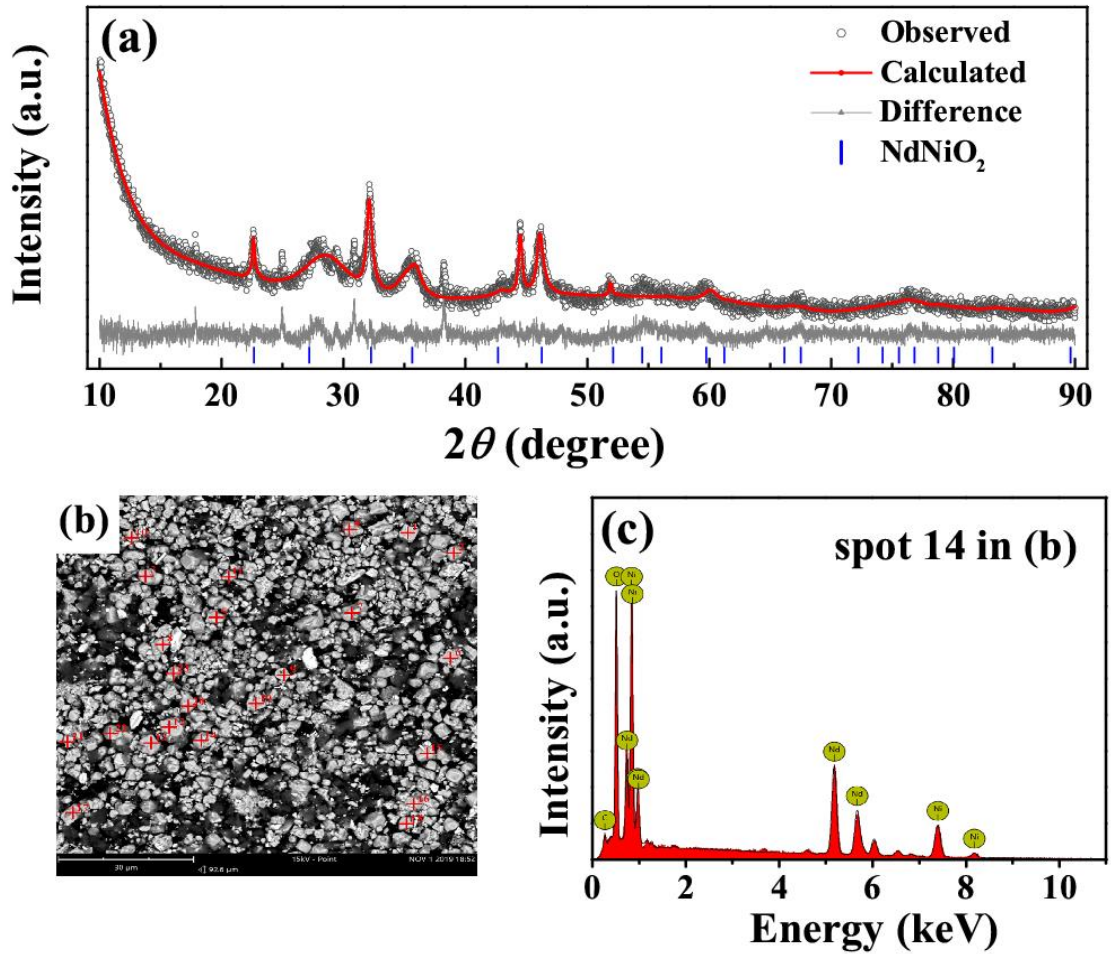

**Supplementary Figure 2 | The Powder X-ray diffraction patterns, SEM image and typical energy dispersive spectroscopy of undoped NdNiO$_2$ sample**. The Rietveld refinements analyses show the formation of infinite-layer 112 phase. The typical EDS pattern of spot 14 (red cross in (b)) confirms the existence of NdNiO$_2$. And the average occupied ratio of nickel in the NiO$_2$ plane of undoped NdNiO$_2$ is 0.91.

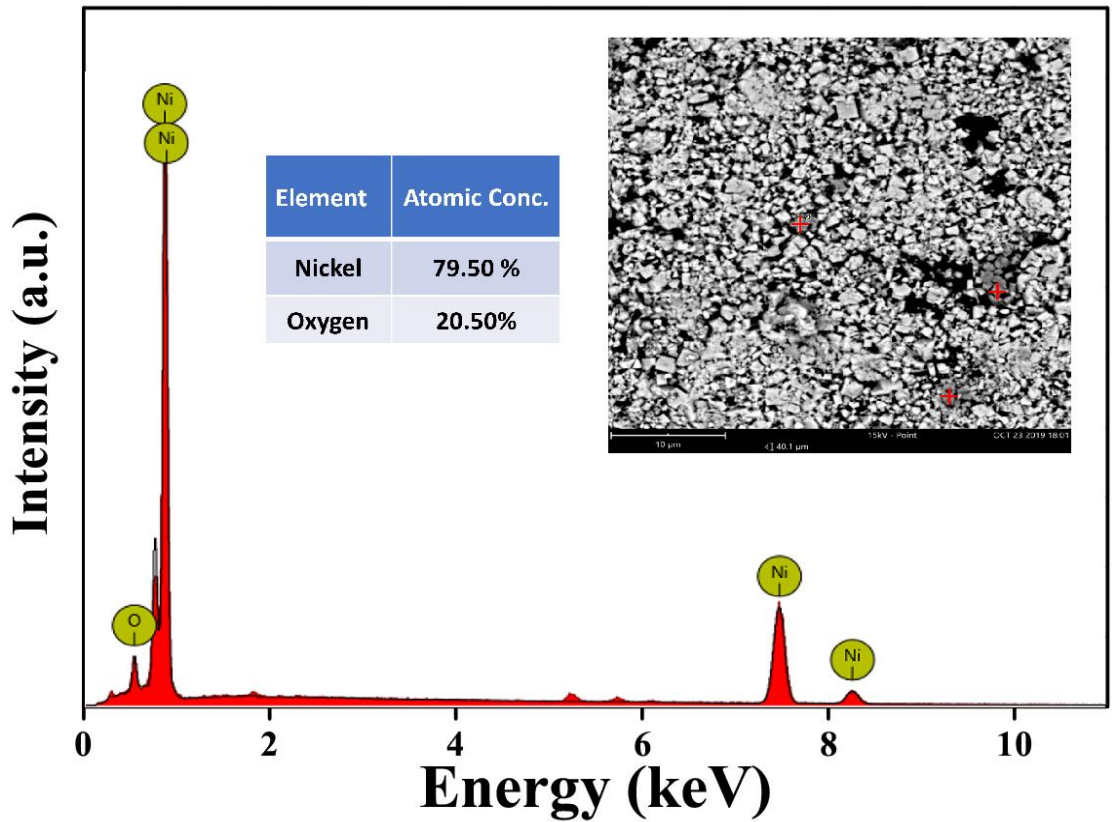

**Supplementary Figure 3 | The typical EDS pattern of the dark grey area in Nd$_{0.8}$Sr$_{0.2}$NiO$_3$ samples**. The energy dispersive X-ray spectroscopy is taken at the red cross of the inset image. The result shows the presence of Ni in our samples.

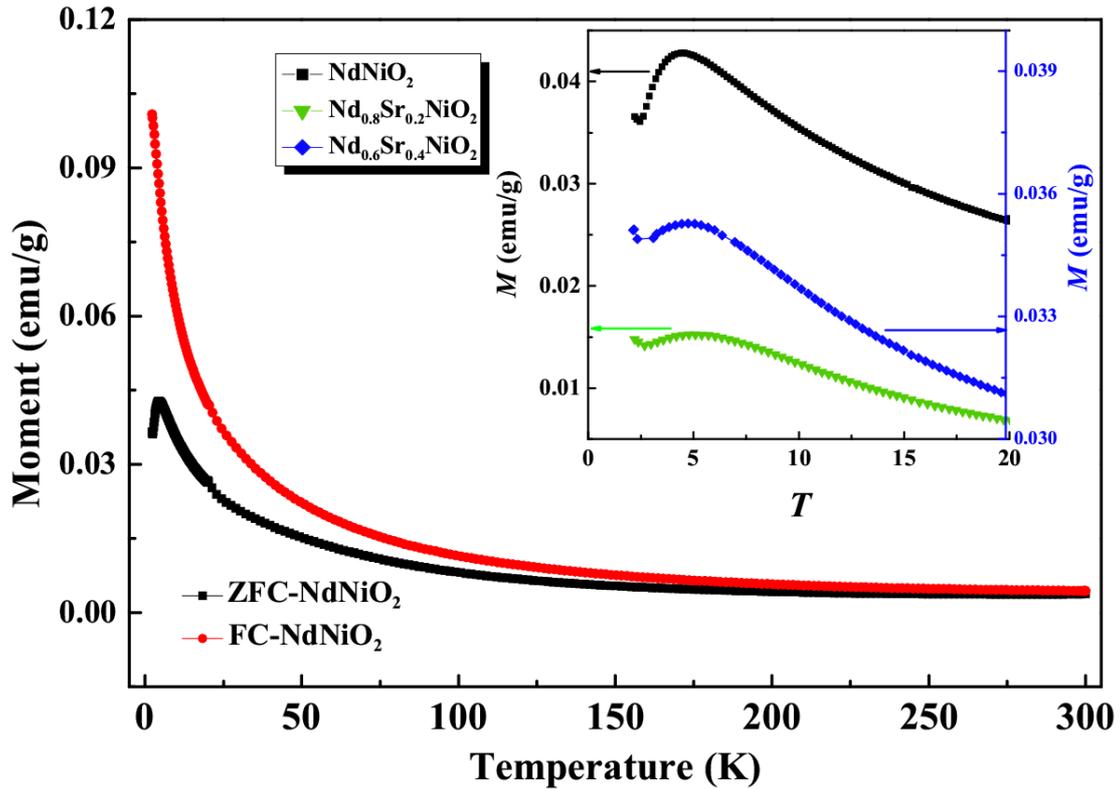

**Supplementary Figure 4 | Magnetic properties of Nd$_{1-x}$Sr$_x$NiO$_2$ ($x$ = 0, 0.2, 0.4) at low field (H = 10 Oe).** Temperature dependence of magnetic moment of NdNiO$_2$ at the field of 10 Oe measured in zero field cooling (ZFC) mode and field cooling (FC) mode. Inset shows the magnetic moment versus temperature of Nd$_{1-x}$Sr$_x$NiO$_2$ (x = 0, 0.2, 0.4) samples with an external field $H$ = 10 Oe in low temperature region with ZFC mode. An obvious irreversibility between the ZFC and FC plots and a maximum magnetization at about 5 K in the ZFC curve are observed.

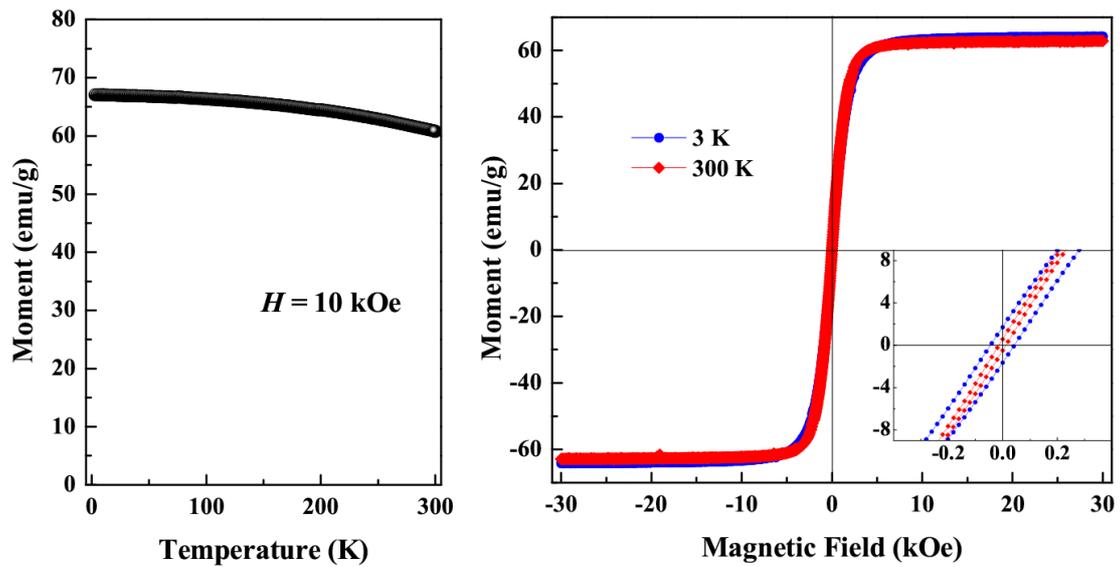

**Supplementary Figure 5 | Magnetic moment versus temperature curve and magnetization hysteresis (*M-H*) loops of pure nickel**. **a,** Temperature dependent magnetic moment of pure nickel at the external magnetic field *H* = 10 kOe. **b,** *M-H* curves of pure nickel at different temperatures with external magnetic fields up to ± 30 kOe.

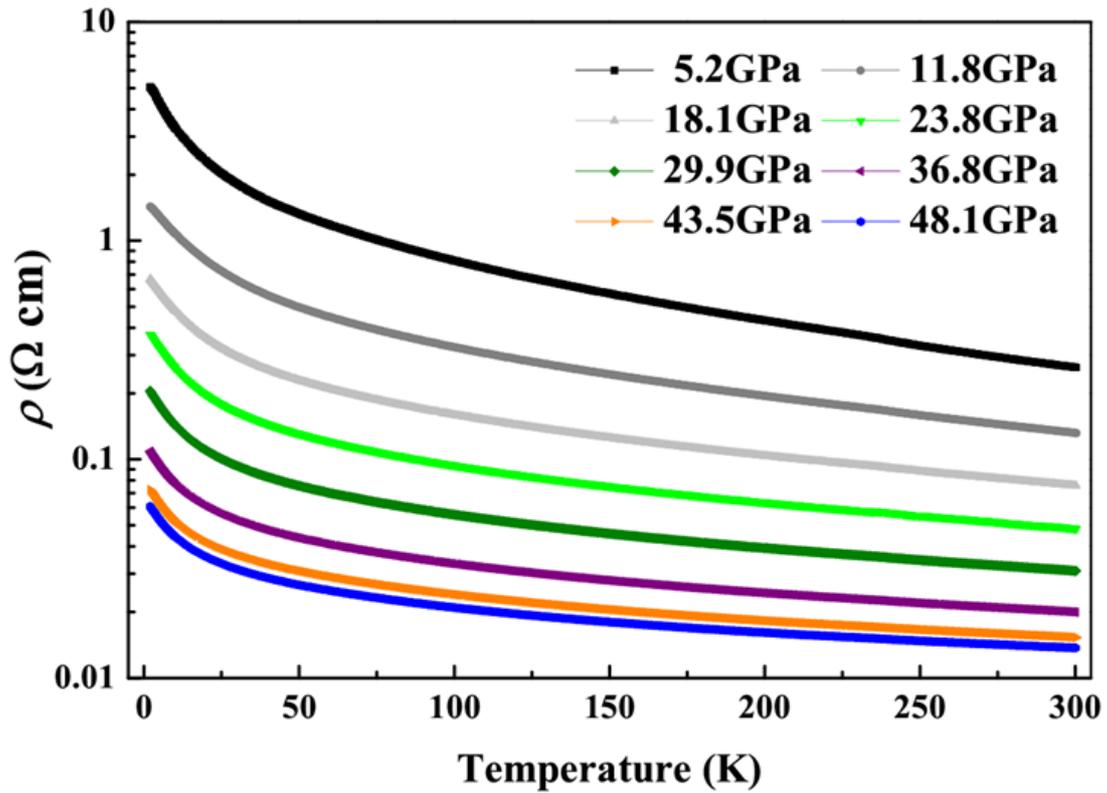

**Supplementary Figure 6 | Temperature dependent resistivity of $Nd_{0.6}Sr_{0.4}NiO_2$ under pressures.** The $\rho_{2K}/\rho_{300K}$ of $Nd_{0.6}Sr_{0.4}NiO_2$ sample is about 20 with the external pressure of 5.2 GPa. With increasing pressure, the magnitude of $\rho_{2K}/\rho_{300K}$ decreases progressively. At the highest pressure of 48.1 GPa, the value of $\rho_{2K}/\rho_{300K}$ has dropped to 4.5, however the $\rho$ (T) curve still exhibits a semiconducting-like feature in the whole temperature region.

## I. XRD characterization of perovskite $Nd_{1-x}Sr_xNiO_3$ (x = 0, 0.2, 0.4) samples

The as-prepared $Nd_{1-x}Sr_xNiO_3$ polycrystalline samples are crushed, ground, washed by distilled water and dried. Then, the XRD patterns are measured at room temperature, as shown in Supplementary Fig. 1. We can see that the perovskite 113 phase is well formed. The rather clean XRD data indicates the high quality of the samples. Two tiny peaks which cannot be indexed with $Nd_{1-x}Sr_xNiO_3$ come from the small amount of unreacted NiO. Besides, the higher Sr-doped concentration would bring in unreacted $Nd_{2-2x}Sr_{2x}NiO_4$ phase in the 113 phase, as shown in the top panel of Supplementary Fig. 1.

## II. The magnetic properties of pure nickel at different fields and temperatures and the Curie-Weiss fitting of effective magnetic moment in $Nd_{1-x}Sr_xNiO_2$ (x = 0.2, 0.4) samples.

As reported previously [S1-S3], the nickel undergoes a transition from paramagnetism to ferromagnetism at about 610 K - 650 K. And the average saturated magnetic moment at low temperature (below 300 K) is about 0.6 $\mu_B$/atom. We also measure the magnetization behavior of nickel at low temperature (see Supplementary Fig. 5), a typical ferromagnetic hysteresis loop with a saturation field about 10 kOe can be seen.

The Curie-Weiss law is used to fit the paramagnetic term of $Nd_{0.8}Sr_{0.2}NiO_2$ and $Nd_{0.6}Sr_{0.4}NiO_2$ samples in the temperature region below 40 K by the equation: $\chi = \chi_0 + \frac{C}{T+T_\theta}$, where $T_\theta$, $\chi_0$, and $C$ can be derived from the fitting parameters. By using the formula $C = \mu_0 \mu_{\text{eff}}^2/3k_B$, one can derive the local magnetic moment $\mu_{eff}$. We obtain the fitting results of Nd$_{0.8}$Sr$_{0.2}$NiO$_2$ ($\chi_0$ = 7.11×10$^{-3}$ emu·mol$^{-1}$Oe$^{-1}$, $T_\theta$ = 4.4 K, and $C$ = 0.674 emu·K·mol$^{-1}$Oe$^{-1}$) and Nd$_{0.6}$Sr$_{0.4}$NiO$_2$ ($\chi_0$ = 7.29×10$^{-3}$ emu·mol$^{-1}$Oe$^{-1}$, $T_\theta$ = 5.2 K, and $C$ = 0.517 emu·K·mol$^{-1}$Oe$^{-1}$) samples. The fitting yields the effective magnetic moments $\mu_{eff}$ of about 2.32 $\mu_B$/f.u. for Nd$_{0.8}$Sr$_{0.2}$NiO$_2$ and 2.03 $\mu_B$/f.u. for Nd$_{0.6}$Sr$_{0.4}$NiO$_2$ samples, respectively.